\documentclass[a4paper,12pt]{article}
\usepackage{amssymb,amsthm,amsfonts,amsmath,cite,enumerate,float}
\usepackage[dvips]{graphicx}
\usepackage[usenames,dvipsnames]{color}
\usepackage[onehalfspacing]{setspace}
\usepackage[labelsep=period]{caption}

\usepackage[symbol*]{footmisc}

\usepackage[pdfencoding=auto]{hyperref}
\usepackage{xcolor}
\usepackage{hyperref}
\definecolor{linkcolor}{HTML}{799B03}
\definecolor{urlcolor}{HTML}{799B03}
\hypersetup{pdfstartview=FitH,linkcolor=linkcolor,urlcolor=urlcolor,colorlinks=true}

\usepackage{geometry}
\begin{document}
\begin{flushright}
INR-TH-2015-019
\end{flushright}
\vspace{10pt}

\begin{center}
{\LARGE \bf Superluminality\\
in dilatationally invariant\\
generalized Galileon theories}\\

\vspace{20pt}

R. S. Kolevatov$^{a,b}$\footnote[2]{\textbf{e-mail:} kolevatov@ms2.inr.ac.ru}

\vspace{15pt}

$^a$\textit{Institute for Nuclear Research of the Russian Academy of Sciences,\\
60th October Anniversary Prospect, 7a, 117312 Moscow, Russia}\\

\vspace{5pt}

$^b$\textit{Department of Particle Physics and Cosmology, Physics Faculty,\\
M.V. Lomonosov Moscow State University,\\
Vorobjevy Gory, 119991, Moscow, Russia}
\end{center}

\vspace{5pt}

\begin{abstract}
We consider small perturbations about homogeneous backgrounds in dilatationally invariant Galileon models. The issues we address are stability (absence of ghosts and gradient instabilities) and superluminality. We show that in the Minkowski background, it is possible to construct the Lagrangian in such a way that any homogeneous Galileon background solution is stable and small perturbations 
about it are subluminal. On the other hand, in the case of FLRW backgrounds, for any Lagrangian functions there exist homogeneous background solutions to the Galileon equation of motion and time dependence of the scale factor, such that the stability conditions are satisfied, but the Galileon perturbations propagate with superluminal speed.
\end{abstract}

\section{Introduction}
Generalized Galileon theories attract considerable attention, mostly because they can violate the null energy condition (NEC) without facing obvious instabilities. The Lagrangians of these theories include second-order derivatives, and yet the equations of motion are second-order differential equations, which is a prerequisite for the absence of ghosts that generically plague higher-derivative theories.

Theories of this sort were first proposed back in 1974 by Horndeski \cite{Horndeski1974}, rediscovered in modified form in Refs. \cite{Fairlie1991,Fairlie1992,Fairlie1992a}, and relatively recently emerged in Refs. \cite{Luty2003,Nicolis2004} in the context of the DGP model \cite{Dvali2000}. Eventually, these theories were considered in their own right \cite{Nicolis2008}. Since then, Galileon theories and their generalizations have been intensely studied in various contexts \cite{Deffayet2009,Chow2009,Deffayet2009a,Deffayet2010,Padilla2010,Kobayashi2010,Hinterbichler2010,Padilla2010a,Goon2010,Deffayet2011,Kobayashi2011,Hinterbichler2012}. Some generalized Galileon theories go under the names of k-Mouflage gravity \cite{Babichev2009}, kinetic gravity braiding \cite{Deffayet2010a,Pujolas2011}, and FabFour models \cite{Charmousis2011,Charmousis2011a}.

Among the uses of the generalized Galileons are the construction of alternatives to inflation, such as the Genesis scenario \cite{Creminelli2010} and bouncing universe models \cite{Qiu2011,Easson2011,Osipov2013,Koehn2013}, and also an attempt to describe the creation of a universe in the laboratory \cite{Rubakov2013}. A review of the theories with the NEC violation is given in Ref. \cite{Rubakov2014}. Many of these applications make use of the Lagrangians, which are invariant in Minkowski space-time under dilatations
\begin{equation}
\label{eq:dilatations}
x^\mu\rightarrow\lambda x^\mu,\quad \pi\rightarrow\pi+\ln\lambda \, .
\end{equation}
The most widely used Lagrangians are
\begin{equation}
\label{eq:Lagrangian}
L=F(Y)e^{4\pi} + K(Y)\square\pi\cdot e^{2\pi} \, ,
\end{equation}
where
\begin{equation}
\begin{aligned}
\label{eq:notations}
&Y = e^{-2\pi}(\partial\pi)^2 \, ,\\
&\square\pi = g^{\mu\nu}\nabla_\mu\nabla_\nu\pi \, ,\\
&(\partial\pi)^2 = g^{\mu\nu}\partial_\mu\pi\partial_\nu\pi \, .
\end{aligned}
\end{equation}

Besides the stability issues, generalized Galileon theories generically face superluminality problem. Superluminal propagation of perturbations about otherwise healthy backgrounds has been argued \cite{Adams2006} (see also Ref. \cite{Camanho2014}) to signal the absence of Lorentz-invariant UV completion (in other words, a model with this property cannot emerge as a low-energy effective theory of some Lorentz-invariant quantum theory, valid at all energy scales). Superluminal propagation has been found in the DGP model \cite{Hinterbichler2009}, in the original Galileon model \cite{Nicolis2009}, and also in a class of bi- and multi-Galileon models \cite{deFromont2013,Garcia-Saenz2013}. Furthermore, original Galileon models admit backgrounds with closed timelike curves \cite{Evslin2011}. In the context of the Genesis scenario, superluminality can be avoided in the vicinity of the relevant background by a judicial choice of the Lagrangian functions $F(Y)$, $K(Y)$ \cite{Creminelli2012,Hinterbichler2012}; however, at least in some cases superluminality is back in the presence of external matter \cite{Easson2013}.

In this paper we address the superluminality issue in dilatationally invariant generalized Galileon theories (\ref{eq:Lagrangian}) in spatially flat FLRW backgrounds
\begin{equation}
\label{eq:FLRW_metric}
\mathrm{d}s^2 = \mathrm{d}t^2 - a^2(t)\mathrm{d}x^2 \, .
\end{equation}
We allow for the NEC violation, and hence impose no restrictions on $H=\dot{a}/a$ and $\dot{H}$. Unlike in previous studies, we do not choose concrete forms of the Lagrangian functions $F(Y)$, $K(Y)$; the question we ask is whether there exist the functions $F(Y)$ and $K(Y)$, such that the perturbations about \textit{any stable} homogeneous background $\pi_c(t)$ are not superluminal. By stability we mean the absence of ghosts and gradient instabilities. Our answer is negative: we find that for any $F(Y)$ and $K(Y)$ there exist values of $H$ and $\dot{H}$ and homogeneous stable solutions to the Galileon field equation, such that the Galileon perturbations about these backgrounds are superluminal. This is the main result of this paper.

We emphasize that our analysis does not rule out the possibility that the superluminality of the Galileon perturbations may be absent in backgrounds obeying the NEC or other energy conditions. Also, we do not claim that the superluminality is actually present in any concrete cosmological model. The reason for choosing FLRW backgrounds is merely technical convenience.

The paper is organized as follows: In Sec. \ref{sec:backgrounds_and_perturbations} we derive the equation of motion for homogeneous Galileons in a spatially flat FLRW metric and quadratic Lagrangian for Galileon perturbations. In Sec. \ref{sec:superluminality_in_FLRW_backgrounds} we obtain our main result: for any choice of $F(Y)$ and $K(Y)$ there exist stable backgrounds about which the perturbations propagate superluminally. We conclude in Sec. \ref{sec:conclusion}.
In the \hyperref[app:Minkowski]{Appendix} we study the theory in Minkowski background and give an explicit example of the Lagrangian functions such that any homogeneous background is stable and perturbations are always subluminal.

\section{Backgrounds and perturbations}
\label{sec:backgrounds_and_perturbations}
The energy-momentum tensor following from Lagrangian (\ref{eq:Lagrangian}) is
\begin{equation}
\label{eq:energy_momentum_tensor}
\begin{split}
&T_{\mu\nu} = 2F'e^{2\pi}\partial_\mu\pi\partial_\nu\pi - g_{\mu\nu}Fe^{4\pi} \\
&+ 2K'\partial_\mu\pi\partial_\nu\pi\square\pi + g_{\mu\nu}\partial_\rho\pi\nabla^\rho(Ke^{2\pi}) - \partial_\mu\pi\nabla_\nu(Ke^{2\pi}) - \partial_\nu\pi\nabla_\mu(Ke^{2\pi}) \, .
\end{split}
\end{equation}

The equation of motion reads
\begin{equation}
\begin{split}
\label{eq:field_eq_arbitrary_metric}
&4Fe^{4\pi} - 6F'e^{2\pi}(\partial\pi)^2 - 2F'e^{2\pi}\square\pi + 4F''(\partial\pi)^4 - 4F''\nabla^\mu\nabla^\nu\pi\partial_\mu\pi\partial_\nu\pi \\
&+4Ke^{2\pi}(\partial\pi)^2 + 4Ke^{2\pi}\square\pi -4K'(\partial\pi)^4 - 4K'(\partial\pi)^2\square\pi - 2K'(\square\pi)^2 + 2K'\nabla^\mu\nabla^\nu\pi\nabla_\mu\nabla_\nu\pi\ \\
&+4K''e^{-2\pi}(\partial\pi)^6 + 4K''e^{-2\pi}(\partial\pi)^4\square\pi - 4K''e^{-2\pi}\nabla^\mu\nabla^\nu\pi\partial_\mu\partial_\nu\pi\square\pi \\
&-8K''e^{-2\pi}\nabla^\mu\nabla^\nu\pi\partial_\mu\pi\partial_\nu\pi(\partial\pi)^2 + 4K''e^{-2\pi}(\nabla^\mu\nabla^\rho\pi)(\nabla_\mu\nabla_\nu\pi)\partial_\rho\pi\partial^\nu\pi \\
&+2K'R_{\mu\nu}\partial^\mu\pi\partial^\nu\pi =0 \, ,
\end{split}
\end{equation}
where the prime denotes a derivative with respect to $Y$. We consider the model in a spatially flat FLRW metric (\ref{eq:FLRW_metric}). We assume that the background solution $\pi=\pi(t)$ is spatially homogeneous. In this case the field equation becomes
\begin{equation}
\label{eq:field_eq_FLRW}
\begin{split}
&4Fe^{4\pi} - 6F'e^{2\pi}\dot{\pi}^2 - 2F'e^{2\pi}\ddot{\pi} + 4F''\dot{\pi}^4 - 4F''\dot{\pi}^2\ddot{\pi} \\
&+ 4Ke^{2\pi}\dot{\pi}^2 + 4Ke^{2\pi}\ddot{\pi} - 4K'\dot{\pi}^4 - 4K'\dot{\pi}^2\ddot{\pi} + 4K''e^{-2\pi}\dot{\pi}^6 - 4K''e^{-2\pi}\dot{\pi}^4\ddot{\pi} \\
&- H(6F'e^{2\pi}\dot{\pi} - 12Ke^{2\pi}\dot{\pi} + 12K'\dot{\pi}^3 + 12K'\dot{\pi}\ddot{\pi} - 12K''e^{-2\pi}\dot{\pi}^5 + 12K''e^{-2\pi}\dot{\pi}^3\ddot{\pi}) \\
&- 18H^2K'\dot{\pi}^2 - 6\dot{H}K'\dot{\pi}^2 = 0 \, .
\end{split}
\end{equation}

Let us consider perturbations $\chi=\chi(\textbf{x},t)$ about a solution $\pi_c=\pi_c(t)$, $\pi=\pi_c+\chi$. We are interested in high momentum and frequencies modes, so we need only the terms proportional to the second derivatives of perturbations in the field equation. Thus, we retain only the terms containing $\square\chi$ and $\nabla^\mu\nabla^\nu\chi$. We also have to consider the second derivatives of the metric, contained in the Ricci tensor \cite{Deffayet2010b}. It follows from Eq. (\ref{eq:energy_momentum_tensor}) that Einstein equation $R_{\mu\nu} - \frac{1}{2}g_{\mu\nu}R = T_{\mu\nu}$ contains the second derivatives of the Galileon [we set $M_{pl}=(8\pi G)^{-\frac{1}{2}} = 1$]. So, one integrates the metric perturbations out from the equation of motion for perturbations by using the Einstein equation. The linearized equation for perturbations reads

\begin{equation}
\label{eq:field_eq_for_perturb}
\begin{split}
&-2F'e^{2\pi}\square\chi - 4F''\partial_\mu\pi\partial_\nu\pi\nabla^\mu\nabla^\nu\chi \\
&+4Ke^{2\pi}\square\chi - 4K'(\partial\pi)^2\square\chi + 4K'\nabla_\mu\nabla_\nu\pi\nabla^\mu\nabla^\nu\chi - 4K'\square\pi\square\chi \\
&+4K''e^{-2\pi}(\partial\pi)^4\square\chi - 4K''e^{-2\pi}\partial_\mu\pi\partial_\nu\pi\square\pi\nabla^\mu\nabla^\nu\chi -  4K''e^{-2\pi}\partial_\mu\pi\partial_\nu\pi\nabla^\mu\nabla^\nu\pi\square\chi \\
&-8K''e^{-2\pi}\partial_\mu\pi\partial_\nu\pi(\partial\pi)^2\nabla^\mu\nabla^\nu\chi + 8K''e^{-2\pi}\partial_\rho\pi\partial^\nu\pi(\nabla_\mu\nabla_\nu\pi)(\nabla^\mu\nabla^\rho\chi) \\
&+ 2K'R^{(1)}_{\mu\nu}\partial^\mu\pi\partial^\nu\pi +  \cdots = 0 \, ,
\end{split}
\end{equation}
where the dots denote terms without second derivatives, and we omit the subscript $c$ of $\pi_c$. $R^{(1)}_{\mu\nu}$ is the linearized Ricci tensor, expressed through the Galileon perturbations:
\begin{equation}
2K'R^{(1)}_{\mu\nu} = 2(K')^2(\partial\pi)^4\square\chi - 8 (K')^2(\partial\pi)^2\partial_\mu\pi\partial_\nu\pi\nabla^\mu\nabla^\nu\chi \, .
\end{equation}
The Lagrangian for perturbations, leading to (\ref{eq:field_eq_for_perturb}), reads
\begin{equation}
\label{eq:Lagrangian_for_perturbations_arbitrary_metric}
\begin{split}
L^{(2)} &= F'e^{2\pi}(\partial\chi)^2 - [2Ke^{2\pi} - \nabla^\mu(K'\nabla_\mu\pi) - K'\square\pi - 2K'(\partial\pi)^2 + (K')^2(\partial\pi)^4](\partial\chi)^2 \\
&+ 2F''\partial_\mu\pi\partial_\nu\pi\partial^\mu\chi\partial^\nu\chi - [2\nabla_\mu(K'\nabla_\nu\pi) - 2K''e^{-2\pi}\partial_\mu\pi\partial_\nu\pi\square\pi \\
&- 4(K')^2(\partial\pi)^2\partial_\mu\pi\partial_\nu\pi]\partial^\mu\chi\partial^\nu\chi + \cdots \, .
\end{split}
\end{equation}
For a homogeneous background we have $\square\pi=\ddot{\pi}+3H\dot{\pi}$, so the Lagrangian for perturbations reads
\begin{equation}
\label{eq:Lagrangian_for_perturbations_FLRW}
L^{(2)}=U\dot{\chi}^2-\dfrac{1}{a^2}V(\partial_i\chi)^2+W\chi^2 \, ,
\end{equation}
where
\begin{subequations}
\label{eq:U&V}
\begin{align}
&U=e^{2\pi}[F' + 2F''Y - 2K + 2K'Y + 2K''Y^2 + 3(K')^2Y^2e^{2\pi}] + 6HK'\dot{\pi} + 6HK''e^{-2\pi}\dot{\pi}^3 \, ,\\
&V=e^{2\pi}[F' - 2K + 2K'Y - 2K''Y^2 - (K')^2Y^2e^{2\pi}] + [ 2K' + 2K''Y]\ddot{\pi} + 4HK'\dot{\pi} \, .
\end{align}
\end{subequations}
We do not need the explicit form of $W$.

The ghosts and gradient instabilities are absent iff
\begin{equation}
\label{eq:U>0&V>0}
U>0 \, ,\qquad V>0 \, .
\end{equation}
Provided these conditions are satisfied, the propagation velocity of perturbations does not exceed the speed of light for
\begin{equation}
\label{eq:U>=V}
\dfrac{V}{U} \leq 1 \, .
\end{equation}
In the \hyperref[app:Minkowski]{Appendix} we consider the Minkowski background metric and give an example of the Lagrangian functions $F(Y)$ and $K(Y)$, such that Eqs. (\ref{eq:U>0&V>0}) and (\ref{eq:U>=V}) are satisfied for all background solutions. So, superluminality can be avoided in the Minkowski background. As we will now see, the situation is entirely different for general FLRW backgrounds.
\section{Superluminality in FLRW backgrounds}
\label{sec:superluminality_in_FLRW_backgrounds}
Our main purpose is to show that if we do not impose any restrictions on the scale factor $a(t)$ (more precisely, on $H$ and $\dot{H}$), 
then there are stable solutions $\pi(t)$ of the field equation (\ref{eq:field_eq_FLRW}), such that the perturbations about them have superluminal speed. This statement is valid for any choice of $F(Y)$ and $K(Y)$ in the Lagrangian (\ref{eq:Lagrangian}).

The coefficients $U$ and $V$ defined by Eq. (\ref{eq:U&V}) are invariant under $T$-transformation:
\begin{equation}
\label{eq:T-transformations}
\dot{\pi}\rightarrow-\dot{\pi},\quad \ddot{\pi}\rightarrow\ddot{\pi},\quad H\rightarrow-H \, .
\end{equation}
Therefore, without loss of generality, we assume that $\dot{\pi}>0$. Let us define the new dimensionless parameters $q$ and $p$ as follows:
\begin{equation}
\label{eq:q&p_parameters}
\begin{aligned}
&H=\dfrac{\dot{a}}{a} = q \cdot \sqrt{Y}e^{\pi} \, ,\\
&\dot{H}=\dfrac{\ddot{a}}{a} - \dfrac{\dot{a}^2}{a^2} = p \cdot Ye^{2\pi} \, .
\end{aligned}
\end{equation}
Note that $q$ and $p$ may have arbitrary signs. We express $\ddot{\pi}$ through $Y$ by making use of the equation of motion (\ref{eq:field_eq_FLRW}):
\begin{equation}
\label{eq:pi_2-dots_FLRW}
\ddot{\pi}=\dfrac{Z'Y + 3q(KY + K'Y^2) - 2Z - 3qF'Y - 6qKY - 9q^2K'Y^2 - 3pK'Y^2}{Z' + 3q(K + K'Y)} \cdot e^{2\pi} \, ,
\end{equation}
where
\begin{equation}
\label{eq:Z_FLRW}
Z=-F + 2F'Y - 2(1 + 3q)KY + 2(1 + 3q)K'Y^2 \, ,
\end{equation}
and $Z'=\dfrac{\partial Z(Y,\pi,H,\dot{H})}{\partial Y}$ (note that $\dfrac{\partial q}{\partial Y}=-\dfrac{q}{2Y}$). In these notations, the functions entering Eq. (\ref{eq:U&V}) are
\begin{subequations}
\label{eq:U&V_FLRW}
\begin{align}
U&=e^{2\pi}[F' + 2F''Y - 2K + 2(1 + 3q)\cdot(K'Y + K''Y^2) + 3(K')^2Y^2e^{2\pi}] \nonumber\\ 
&=e^{2\pi}[Z' + 3q(K + K'Y) + 3(K')^2Y^2e^{2\pi}]  \, \label{eq:U_FLRW} \\	
V&=e^{2\pi}[F' - 2K + 4(1 + q)K'Y - 2(K' + K''Y)\dfrac{2Z + 3qF'Y + 6qKY + 9q^2K'Y^2 + 3pK'Y^2}{Z' + 3q(K + K'Y)} \nonumber\\
&- (K')^2Y^2e^{2\pi}] \, .\label{eq:V_FLRW}
\end{align}
\end{subequations}
It is sufficient for our purposes to consider sufficiently large negative fields $\pi$, so we can omit the terms $3(K')^2Y^2e^{2\pi}$ and $(K')^2Y^2e^{2\pi}$ in (\ref{eq:U_FLRW}) and (\ref{eq:V_FLRW}), respectively. Note that after omitting these terms the expression in the denominator in Eq. (\ref{eq:V_FLRW}) is always positive in the region of stability, because this denominator equals $e^{-2\pi}U$.

The values of $Y$, $q$ and $p$ in Eq. (\ref{eq:U&V_FLRW}) are arbitrary,
except for the restriction
$Y>0$. The fact that the background $\pi$ obeys the equation of motion has been used
to remove $\ddot{\pi}$ from the expression for $V$.

\subsection{Regions of stability and regions of subluminal speed}
The stability conditions $U>0$, $V>0$ and the condition of the absence of superluminality $U>V$ can be viewed as inequalities for polynomials in $q$ and $p$ with coefficients depending on $Y$. The inequality $U>0$ is equivalent to the condition
\begin{equation}
\label{eq:U>0_FLRW}
U>0: \qquad F' + 2F''Y - 2K + 2M(1 + 3q) > 0 \, ,
\end{equation}
where 
\begin{equation}
M=K'Y + K''Y^2.
\end{equation}
The inequality $V>0$ is equivalent to
\begin{equation}
\label{eq:V>0_FLRW}
V>0: \qquad C_1 + C_2q + C_3q^2 - C_4p > 0 \, ,
\end{equation}
where
\begin{subequations}
\label{eq:coeffs_for_V}
\begin{align}
&C_1 = (F' - 2K + 4K'Y)(F' + 2F''Y - 2K + 2M) - 8M(F' - K + K'Y) + 4F\dfrac{M}{Y} \, ,\\
&C_2 = 4K'Y(F' + 2F''Y - 2K + 2M) \, ,\\
&C_3 = 6MK'Y \, ,\\
&C_4 = 6MK'Y \, .
\end{align}
\end{subequations}
Note that $C_3=C_4$. The inequalities (\ref{eq:U>0_FLRW}) and (\ref{eq:V>0_FLRW}) determine the domain of stability of the background solution $\pi(t)$. We now ask whether perturbations are not superluminal everywhere in this domain. The condition for the absence of superluminality $U-V\geq 0$ can be written as follows:
\begin{equation}
\label{eq:U>=V_FLRW}
U - V \geq 0: \qquad \tilde{C_1} + \tilde{C_2}q + \tilde{C_3}q^2 + \tilde{C_4}p \geq 0 \, ,
\end{equation}
where
\begin{subequations}
\label{eq:coeffs_for_U-V}
\begin{align}
&\tilde{C_1} = (2F''Y - 4K'Y + 2M)(F' + 2F''Y - 2K + 2M) + 8M(F' - K + K'Y) - 4F\dfrac{M}{Y} \, ,\\
&\tilde{C_2} = 2(6M - 2K'Y)(F' + 2F''Y - 2K + 2M) \, ,\\
&\tilde{C_3} = 36M^2 - 6MK'Y = 36M^2 - C_3 \, , \label{eq:C3&tilde_C3}\\
&\tilde{C_4} = 6MK'Y \, .
\end{align}
\end{subequations}
Note that $\tilde{C_4}=C_4$. 

Let us point out that a theory with identically vanishing $K'$ is trivial: upon integration by parts the second term in the Lagrangian (\ref{eq:Lagrangian}) is reduced to a term without second derivatives. In what follows we assume that
$K' \neq 0$ at least for some $Y$.

Our strategy will be as follows: we consider a given value of $Y$ such that $K' \neq 0$ and ask whether or not there exists a region in the $(q,p)$ plane, such that the background is stable, $U>0$, $V>0$, but perturbations propagate superluminally, $U - V < 0$. The cases $M \neq 0$ and $M = 0$ are to be treated separately.

\subsection{\texorpdfstring{$M \neq 0$}{M does not equal 0}}
Let us first consider the case $M \neq 0$. The domains of stability and regions of subluminality, determined by the conditions (\ref{eq:U>0_FLRW}), (\ref{eq:V>0_FLRW}) and (\ref{eq:U>=V_FLRW}), occupy certain regions in the $(q,p)$ plane. 

From Eq. (\ref{eq:U&V_FLRW}) it follows that the line $U=0$ is determined by the equation
\begin{equation}
\label{eq:line_U=0}
q=-\dfrac{1}{6M}(F' + 2F''Y - 2K + 2M) \, .
\end{equation}
Depending on the sign of $M$, the region (\ref{eq:U>0_FLRW}) is on the right or left of the line (\ref{eq:line_U=0}) in the $(q,p)$ plane.

The boundaries of the regions (\ref{eq:V>0_FLRW}) and (\ref{eq:U>=V_FLRW}) are, respectively, the parabolas
\begin{subequations}
\label{eq:parabolas_V=0&U=V}
\begin{align}
&V=0: \qquad C_4p = C_3q^2 + C_2q + C_1 \, , \label{eq:parabola_V=0} \\
&U=V: \qquad C_4p = -\tilde{C_3}q^2  -\tilde{C_2}q - \tilde{C_1} \, \label{eq:parabola_U=V}. 
\end{align}
\end{subequations}
Parabolas (\ref{eq:parabola_V=0}) and (\ref{eq:parabola_U=V}) are tangent to each other at one point. Indeed, the common point of these parabolas is at $U=V=0$, hence the value of $q$ at this point is given by Eq. (\ref{eq:line_U=0}), while the value of $p$ is determined by Eq. (\ref{eq:parabolas_V=0&U=V}). Furthermore, since, according to Eq. (\ref{eq:C3&tilde_C3}), $-\tilde{C_3} = C_3 - 36M^2 \neq C_3$ (recall that we consider the case $M \neq 0$), the fact that the common point of the parabolas (\ref{eq:parabola_V=0}) and (\ref{eq:parabola_U=V}) is unique implies that these parabolas are tangent to each other at this point.

Due to the relation $C_3+\tilde{C_3}=36M^2$, the coefficients $C_3$ and $\tilde{C_3}$ cannot be negative simultaneously. If $C_3$ and $\tilde{C_3}$ have different signs, then there are two possibilities:
\begin{enumerate}
\item[1]) $\tilde{C_3}<0$: then $C_3=36M^2+\vert\tilde{C_3}\vert$, and the parabola (\ref{eq:parabola_V=0}) increases or decreases faster than the parabola (\ref{eq:parabola_U=V}), depending on the sign of $C_4$.
\item[2]) $C_3<0$: then $\tilde{C_3}=36M^2+\vert C_3\vert$, and the parabola (\ref{eq:parabola_U=V}) increases or decreases faster than the parabola (\ref{eq:parabola_V=0}), depending on the sign of $C_4$.
\end{enumerate}

We recall that $C_3=C_4$ and arrive at three possibilities for the location of the regions of stability and of subluminality in the $(q,p)$ plane, shown in Figs. \ref{fig:image1}-\ref{fig:image3}.

\begingroup
\centering
\begin{figure}[H]
\center{\includegraphics[scale = 0.55]{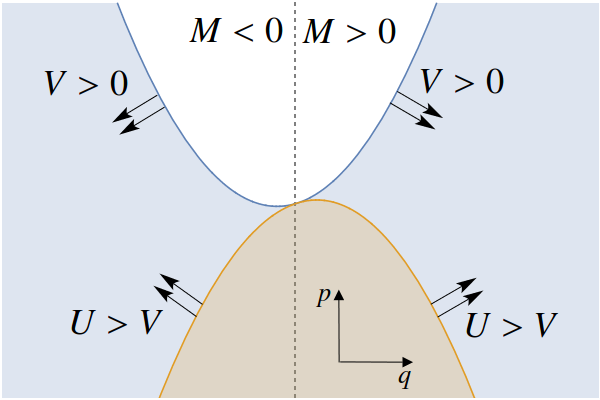}} 
\caption{$C_3>0$, $\tilde{C_3}>0$, $C_4>0$}
\label{fig:image1}
\end{figure}
\nopagebreak
\begin{figure}[H]
\center{\includegraphics[scale = 0.55]{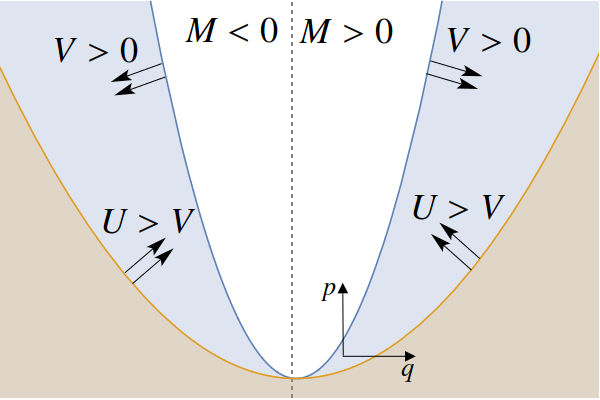}}
\caption{ $C_3>0$, $\tilde{C_3}<0$, $C_4>0$}
\label{fig:image2}
\end{figure}
\begin{figure}[H]
\center{\includegraphics[scale = 0.55]{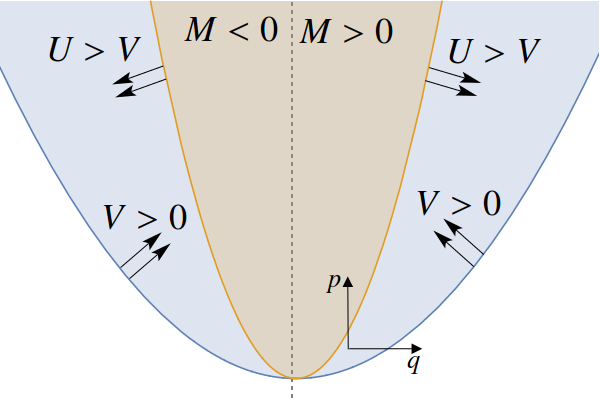}}
\caption{$C_3<0$, $\tilde{C_3}>0$, $C_4<0$}
\label{fig:image3}
\end{figure}
\endgroup

Regions of $V>0$ and $U>V$ are located on the sides of the corresponding parabolas shown by arrows. The region where $U>0$ is to the right (left) of the dashed line for $M>0$ ($M<0$). Backgrounds are stable and there is no superluminal propagation in the blue (light grey) regions (to the right and left of the dashed line for $M>0$ and $M<0$, respectively), while
red (dark grey) regions are such that ghosts and gradient instabilities are absent, but perturbations propagate superluminally. From Figs. \ref{fig:image1}-\ref{fig:image3} it immediately follows that the conditions of stability $U>0$ and $V>0$ do not exclude superluminal propagation $U<V$ if we do not impose any restrictions on $q$ and $p$ (in other words, on $H$ and $\dot{H}$).

\subsection{\texorpdfstring{$M=0$}{M equals 0}}
Let us consider the case $M=0$, when $C_3=\tilde{C_3}=C_4=\tilde{C_4}=0$, and conditions (\ref{eq:U>0_FLRW}), (\ref{eq:V>0_FLRW}) and (\ref{eq:U>=V_FLRW}) do not involve $p$. In this case the stability conditions have the form
\begin{subequations}
\label{eq:U>0&V>0_M=0}
\begin{align}
&U>0: \qquad F' + 2F''Y - 2K > 0 \, ,\label{eq:U>0_M=0} \\
&V>0: \qquad F' - 2K + 4(1 + q)K'Y > 0 \, . \label{eq:V>0_M=0}
\end{align}
\end{subequations}
There is no superluminal propagation iff
\begin{equation}
U - V \geq 0: \qquad 2F''Y - 4(1 + q)K'Y \geq 0 \, .
\end{equation}

Since Eq. (\ref{eq:U>0_M=0}) does not involve $q$, for any given $Y$ obeying (\ref{eq:U>0_M=0}) one can choose large enough $\vert q \vert$ (with
the sign of $q$  determined by the requirement $q \cdot K' > 0$) and obtain $V > 0$ and $U - V < 0$. Hence, there is superluminal propagation about stable backgrounds in the case $M=0$ as well.

\section{Conclusion}
\label{sec:conclusion}

A few remarks are in order. First, the backgrounds with superluminal propagation of the Galileon perturbations are time dependent, so one may wonder whether there is enough time for the superluminality to show up. This is not an issue, however, as the superluminal perturbations have arbitrarily high frequencies (and momenta).

Second, the background metric must be sub-Planckian:
\begin{equation}
\label{eq:sub-Planckian_metric}
\begin{aligned}
&H \ll M_{pl} \, ,\\
&\dot{H} \ll M_{pl}^2 \, .
\end{aligned}
\end{equation}
This is not an issue either. According to our definition  (\ref{eq:q&p_parameters}) of the parameters $q$ and $p$, the conditions (\ref{eq:sub-Planckian_metric}) read
\begin{equation}
\begin{aligned}
\label{eq:sub-Planckian_q&p}
&q \cdot \sqrt{Y}e^{\pi} \ll M_{pl} \, ,\\
&p \cdot Ye^{2\pi} \ll M_{pl}^2 \, .
\end{aligned}
\end{equation}
The stability and subluminality conditions involve $q$, $p$ and $Y$,
but not $\dot{\pi}$ and $\pi$ separately. So,
by reducing $e^\pi$ at fixed $Y$, $q$ and $p$, one can always make the relevant
values of $H$ and $\dot{H}$ arbitrarily small and hence satisfy Eq. (\ref{eq:sub-Planckian_metric}).

To summarize, we have considered the dilatationally invariant Galileon model with the Lagrangian (\ref{eq:Lagrangian}). In the case of Minkowski space-time there exist Lagrangian functions $F(Y)$ and $K(Y)$, for which all homogeneous solutions are stable, and perturbations about them propagate with subluminal speed (see the \hyperref[app:Minkowski]{Appendix}). On the other hand, we have shown that in the case of FLRW backgrounds, for any $F(Y)$ and $K(Y)$ there exist homogeneous background solutions to the Galileon equation of motion and the time dependence of the scale factor, such that the stability conditions are satisfied, but the Galileon perturbations propagate with superluminal speed.

\section*{Acknowledgments}
The author thanks Valery Rubakov for valuable comments and 
Petr Satunin, Dmitry Eremeev, Alan Kanapin, Ivan Markin, Alexey Pustynnikov and Sergey Shirobokov for useful discussions. This work has been supported by Russian Science Foundation grant 14-22-00161.

\section*{Appendix. Minkowski background}
\label{app:Minkowski}
Let us consider the model (\ref{eq:Lagrangian}) in Minkowski background. Our main purpose is to find the
Lagrangian functions $F(Y)$ and $K(Y)$, such that any homogeneous background solution $\pi(t)$ is stable and perturbations about it are subluminal. This means that $U>V>0$ for any homogeneous solution to the Galileon field equation. In the case $a=1$, the coefficients $U$ and $V$ are given by
\begin{subequations}
\label{eq:U&V_Minkowski}
\begin{align}
&U=e^{2\pi}[F' + 2F''Y - 2K + 2K'Y + 2K''Y^2] \, ,\\
&V=e^{2\pi}[F' - 2K + 2K'Y - 2K''Y^2] + [2K' + 2K''Y]\ddot{\pi} \, .
\end{align}
\end{subequations}
The expression for $\ddot{\pi}$ follows from the equation of motion (\ref{eq:field_eq_FLRW}):
\begin{equation}
\label{eq:pi_2-dots_Minkowski}
\ddot{\pi}=\dfrac{Z'Y - 2Z}{Z'} \cdot e^{2\pi} \, ,
\end{equation}
where
\begin{equation}
\label{eq:Z_Minkowski}
Z=-F + 2F'Y - 2KY + 2K'Y^2 \, .
\end{equation}
Hence, the stability conditions $U>0$ and $V>0$ read
\begin{subequations}
\label{eq:U>0&V>0_Minkowski}
\begin{align}
&Z' > 0 \, , \label{eq:U>0_Minkowski} \\
&F' - 2K + 4K'Y - 4[K' + K''Y]\cdot\dfrac{Z}{Z'} > 0 \, . \label{eq:V>0_Minkowski}
\end{align}
\end{subequations}
Let us give an explicit example of the functions $F(Y)$ and $K(Y)$, for which inequalities (\ref{eq:U>0&V>0_Minkowski}) are satisfied for any $Y$. We choose $F(Y)$ as a second-order polynomial in $Y$ and assume that $K(Y)$ depends on $Y$ linearly, i.e.
\begin{equation}
\label{eq:F(Y)&K(Y)}
\begin{aligned}
&F(Y) = aY^2 + bY \, ,\\
&K(Y) = cY \, ,
\end{aligned}
\end{equation}
where $a$ and $b$ are  constant coefficients.
Taking into account Eqs. (\ref{eq:Z_Minkowski}) and  (\ref{eq:F(Y)&K(Y)}), the inequality (\ref{eq:U>0_Minkowski}) has the form
\begin{equation}
\label{eq:U(a,b,c)>0_Minkowski}
6aY + b > 0 \, .
\end{equation}
Recall that $Y=e^{-2\pi}\dot{\pi}^2$ is always positive. The condition (\ref{eq:U(a,b,c)>0_Minkowski}) is satisfied for all $Y$ when $a>0$ and $b>0$.
The inequality (\ref{eq:V>0_Minkowski}) has the form
\begin{equation}
\label{eq:V(a,b,c)>0_Minkowski}
12a^2Y^2 + (8ab - 2bc)Y + b^2 > 0 \, .
\end{equation}
This condition is satisfied for all $Y$, provided that
\begin{equation}
\label{eq:V(a,b,c)>0_Minkowski_forall_Y}
(4a - c)^2 - 12a^2 < 0 \, .
\end{equation}
So, any background solution is stable, provided that
\begin{equation}
\label{eq:U(a,b,c)>0&V(a,b,c)>0_Minkowski}
\begin{cases}
a > 0 \, ,
\\
b > 0 \, ,
\\
4 - 2 \sqrt{3} < \dfrac{c}{a} < 4 + 2 \sqrt{3} \, .
\end{cases}
\end{equation}
The condition of the absence of superluminality is $U\geq V$, which in the notation (\ref{eq:Z_Minkowski}) has the form
\begin{equation}
\label{eq:U>V_Minkowski}
Z'\geq F' - 2K + 4K'Y - 4\cdot[K' + K''Y]\cdot\dfrac{Z}{Z'} \, .
\end{equation}
For functions (\ref{eq:F(Y)&K(Y)}), this reads
\begin{equation}
\label{eq:U(a,b,c)>V(a,b,c)_Minkowski}
12a^2Y + 2ab + bc \geq 0 \, .
\end{equation}
This inequality is always satisfied for the choice of parameters (\ref{eq:U(a,b,c)>0&V(a,b,c)>0_Minkowski}). Note that the inequality (\ref{eq:U(a,b,c)>V(a,b,c)_Minkowski}) is valid in the strict sense, i.e. the speed of perturbations does not reach the speed of light.

Thus, we have shown that all solutions are stable in the Minkowski background, and there are no superluminal perturbations if we choose functions $ F(Y)$ and $K(Y)$ in the Lagrangian (\ref{eq:Lagrangian}) in the form of (\ref{eq:F(Y)&K(Y)}) with the constant coefficients $a$, $b$ and $c$ in the range (\ref{eq:U(a,b,c)>0&V(a,b,c)>0_Minkowski}).

\end{document}